\documentclass[aps,prl,epsfig,floats,twocolumn,superscriptaddress,amssymb,amsmath,floatfix,showpacs]{revtex4}
\usepackage{graphicx}
\begin{document}

\title{Propulsion of a molecular machine by asymmetric
  distribution of reaction--products}

\author{Ramin Golestanian}

\affiliation{Isaac Newton Institute for Mathematical Sciences,
Cambridge, CB3 0EH, UK}

\affiliation{Institute for Advanced Studies in Basic Sciences,
Zanjan 45195-159, Iran}

\affiliation{Department of Applied Mathematics, University of
Leeds, Leeds LS2 9JT, UK}

\author{Tanniemola B. Liverpool}

\affiliation{Isaac Newton Institute for Mathematical Sciences,
Cambridge, CB3 0EH, UK}

\affiliation{Department of Applied Mathematics, University of
Leeds, Leeds LS2 9JT, UK}

\author{Armand Ajdari}

\affiliation{Laboratoire de Physico-Chimie Th\'eorique, UMR CNRS
7083, ESPCI, 10 rue Vauquelin, F-75231 Paris Cedex 05, France}

\date{\today}

\begin{abstract}
  A simple model for the reaction-driven propulsion of a small device
  is proposed as a model for (part of) a molecular machine in aqueous
  media. Motion of the device is driven by an asymmetric
  distribution of reaction products. The propulsive 
  velocity of the device is calculated as well as the scale of the
  velocity fluctuations. The effects of hydrodynamic flow as well as a
  number of different scenarios for the kinetics of the reaction are
  addressed.
\end{abstract}

\maketitle


Molecular motors are machines that convert chemical energy to
mechanical work~\cite{motors,armand}. Examples are the cytoplasmic
motors that move along biological (protein) tracks in the cell by
converting the energy released upon ATP hydrolysis into mechanical
work~\cite{howard,mackay}. These complex machines act as the
inspiration for the design of macromolecular
devices~\cite{molmach} with the ability to sort, sense and
transport material in chip-sized laboratories~\cite{soong}.
Consequently a major area of current chemical research is the
construction of much simpler {\em molecular machines} for
nanotechnological applications~\cite{leigh}.

In this spirit, we study in this Letter a simple model of a
self-propelling device driven by chemical reactions on its
surface. It is simple enough that not only is the construction
using chemical techniques feasible, but it should also be
possible to change parameters in order to optimize particular
features  or functions.

\begin{figure}
\includegraphics[width=.6\columnwidth]{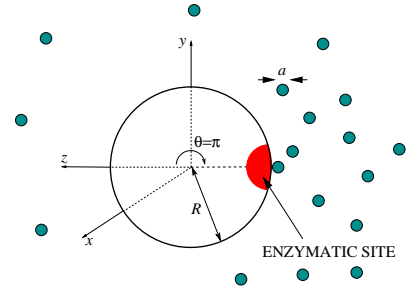}
\caption{The spherical particle with a reaction site.
} \label{fig:schem}
\end{figure}

We consider a spherical particle (colloid or vesicle) of radius
$R$ that has a {\em single} enzymatic site located on its surface
at a fixed position, as sketched in Fig. \ref{fig:schem}. In the
presence of a reactive substrate in a non-equilibrium state, the
enzyme promotes the reaction rate in its vicinity and produces a
dynamic and asymmetric distribution of product particles of
(hydrodynamic) diameter $a \ll R$ which exert osmotic or
interfacial forces (depending the boundary properties) and hence
propel the sphere in a fixed direction. We consider both uniform
and periodic reactions and calculate the propulsive velocity of
the sphere. We find that the velocity of propulsion is set by the
size of the product particles, the properties on the boundary, and
the reaction rate. Upon considering the variations in the rate
particle release (and taking account of density fluctuations), we
find that the ratio of the mean-square velocity fluctuations to
the mean velocity depends on the ratio of the time for a product
particle to diffuse a distance $R$ to the typical time between the
production of successive product particles.

The reaction site on the sphere, located at ${\bf r}_s=-{\bf {\hat
z}} R$ (see Fig. \ref{fig:schem}), is an enzyme catalyzing the
break up of an available substrate into two product particles. If
one of the product particles is similar in size to the original
substrate (see below), we can assume that the reaction site
effectively acts as a source of excess particles that are being
released at the reaction site at a rate $\frac{d N_p(t)}{d t}$,
and the diffusion equation for the density of the excess product
particles can be written as
\begin{equation}
\partial_t \rho({\bf r},t)-D \nabla^2 \rho({\bf r},t)=\frac{d N_p(t)}{d t}
\;\delta^3({\bf r}-{\bf r}_s),\label{diff-eq-1}
\end{equation}
where $D$ is the diffusion constant of these particles. The
density profile can be obtained from Eq. (\ref{diff-eq-1})
subject to the boundary condition of vanishing normal current on
the surface of the sphere, ${\bf {\hat r}} \cdot \nabla \rho|_{\rm
sphere}=0$.

The resulting distribution of product particles around the sphere is
asymmetric, with a non-zero first moment of $\rho_1(t)=\int_0^\pi
\sin\theta d \theta \; \cos \theta \; \rho(r=R,\theta,t)$, leading to
phoretic motion of the sphere.  It is well known that colloidal
particles in {\em externally} imposed solute gradients will be set in
motion by a variety of phoretic mechanisms \cite{Anderson-review}.
Here in contrast, the gradient (of products) is {\em
  self-generated}~\cite{Lammert} by the device itself.
We thus obtain a general expression for the
velocity of propulsion in the $z$-direction
\begin{math}
v=-\frac{k_{\rm B} T}{\eta}\; \frac{\lambda^2}{R}\;
\rho_1,
\end{math}
where $\eta$ is the viscosity of the solvent and the length-scale
$\lambda$ depends on the particular phoretic
mechanism\cite{Anderson-review,ToBePub}.  The other components of the
velocity vanish due to symmetry.  Two mechanisms valid for
non-ionic product particles are the ``diffusiophoresis'' of a totally
impermeable sphere \cite{Derjaguin}, because the depletion of the
product particles near its surface causes a lateral slip velocity that
results in net motion of the sphere, and the ``osmiophoresis'' of a
spherical shell which is permeable to solvent but impermeable to
product particles which develops a non-zero velocity due to osmotic
forces that cause radial flows of solvent across the
membrane\cite{Anderson-osmio}.

Experimental estimates for the diffusiophoretic, $\lambda_D$ and the
osmiophoretic, $\lambda_O$ lengthscales are available from the
literature.  Measurements on latex particles in gradients of Dextran
by Staffeld and Quinn \cite{Staffeld} obtained values of $\lambda_D
\simeq 38 \, \mbox{nm}$.  Studies of micron sized lipid vesicles of
radius $R$ in gradients of sucrose by Nardi et al ~\cite{Nardi} have
obtained an osmiophoretic lengthscale $\lambda_O \sim 0.3 R $.
Theoretical treatments of phoretic propulsion implement local momentum
conservation by solving the Stokes' equation for the solvent (Re$\ll
1$) taking account of the product particles and force balance on the
sphere \cite{Anderson-review}.  For diffusiophoresis, Anderson and
Prieve~\cite{Anderson-Prieve} have calculated
\begin{math}
\lambda^2_D=\int_{0}^\infty d l \;l \left[1-{\rm
e}^{-{W(l)}/{k_{\rm B} T}}\right],
\end{math}
where $W(l)$ is the interaction energy between the solute particles
and the rigid wall of the sphere at a normal separation $l$
\cite{Derjaguin}. This is consistent with the experiments above \cite{Staffeld}
for an interaction range given by the hydrodynamic radius of Dextran.
For osmiophoresis, Anderson~\cite{Anderson-osmio} has obtained,
\begin{math}
\lambda^2_O=R^2 \left[\frac{\eta L_p/R}{2+20 \;\eta L_p/R}\right],
\end{math}
where $L_p$ is the filtration coefficient of the membrane. For a
$R=2\mu$m lipid vesicle, using $\eta=10^{-3} \;{\rm Pa.s}$ and
$L_p=10^{-7} \;{\rm (m/s)/atm}$ \cite{Lp}, one obtains
$\frac{\lambda_O}{R} \sim 0.01$ significantly lower than the
experiments above~\cite{Nardi}. This discrepancy between theory and
experiment remains an open problem \cite{ToBePub}.

Solving Eq. (\ref{diff-eq-1}) subject to the appropriate boundary
condition mentioned above, we find $\rho_1$, which together with
the above expression for the velocity of propulsion yields
\begin{equation}
v(t)=\ell \int \frac{d \omega}{2 \pi} \frac{e^{-i \omega t}
f(\omega) \left[1-i \left(\sqrt{\frac{i \omega
R^2}{D}}\right)\right]}{\left[1-i \left(\sqrt{\frac{i \omega
R^2}{D}}\right)-\frac{1}{2} \left(\sqrt{\frac{i \omega
R^2}{D}}\right)^2\right]}, \label{force-2}
\end{equation}
where $\ell=\frac{k_{\rm B} T}{4 \pi \eta
D}\;\frac{\lambda^2}{R^2}$, and $f(\omega)$ is the Fourier
transform of $\frac{d N_p(t)}{d t}$. Using Einstein's formula for
the diffusion coefficient $D=k_{\rm B} T/(3 \pi \eta a)$, we find
\begin{equation}
\ell=a \; \left(\frac{3 \; \lambda^2}{4 \;
R^2}\right).\label{ell-def}
\end{equation}
The time dependence of the velocity will depend on the particle
release rate. Two important cases of uniform and periodic particle
release are considered.
\paragraph{Uniform Particle Release.}       \label{sec:random}

\begin{figure}
\includegraphics[width=.6\columnwidth]{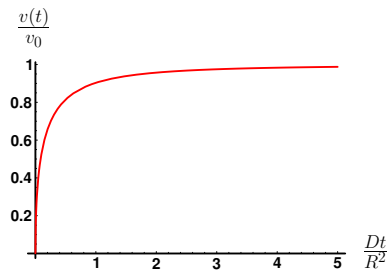}
\caption{Transient response of the velocity of propulsion after the
reaction has been switched on.} \label{fig:force-ini}
\end{figure}

Consider the case where a particle source with a uniform rate is
switched on at $t=0$; i.e. $\frac{d N_p(t)}{d
t}=\frac{1}{\tau_{\rm f}}\vartheta(t)$, where $\tau_{\rm f}$ is
the average inverse reaction rate or ``firing time'' of the
product particles, and $\vartheta(t)$ is the Heaviside step
function. Using Eq. (\ref{force-2}), we find the average velocity
of the sphere as
\begin{math}
v(t)=v_0 \; \vartheta(t) \; {\cal G}\left(\frac{D
t}{R^2}\right),
\end{math}
where the stationary velocity is given by
\begin{equation}
v_0=\frac{\ell}{\tau_{\rm f}},\label{v0-1}
\end{equation}
and ${\cal G}(s)=1-\cos 2 s-\sin 2 s+4 \sqrt{\frac{s}{\pi}}\;
{}_1F_2\left(1;\frac{3}{4},\frac{5}{4};-s^2\right), $ where
${}_1F_2\left(1;\frac{3}{4},\frac{5}{4};-s^2\right)$ is a
generalized hypergeometric function \cite{Hyper}. The time
response of the force is plotted in Fig. \ref{fig:force-ini}. The
function starts at $t=0$ with an infinite slope and asymptotes to
its final value around $t=R^2/D$.

We can estimate the steady state propulsive velocity for
a spherical device of radius $R=2\mu$m   
with the enzyme on its' surface catalyzing a fast reaction such as
that of {\em acetylcholinesterase}, which hydrolyzes 
acetylcholine in the synaptic cleft into components acetate and
choline at the rate of $1/\tau_{\rm f} \simeq 25000 \;{\rm s}^{-1}$
\cite{Stryer}. The corresponding product particle will be acetate with
a hydrated diameter of $a=0.8 \;{\rm nm}$.  Using the estimates
above~\cite{Staffeld,Anderson-Prieve} for $\lambda_D$ we find a
steady-state propulsive velocity from diffusiophoresis of a hard
sphere as $v_{\rm D} \sim 1\;{\rm n m}/{\rm s}$. This rather small
value could be improved by ``engineering'' the strength and range of
the interaction of products with the surface.  The high experimental
value of $\lambda_O$ measured \cite{Nardi} for closed vesicles imply a
much higher velocity of propulsion $v_{\rm O} \sim 1\; \mu{\rm m}/{\rm
  s}$ for a semipermeable shell.  Theoretical estimates
\cite{Anderson-osmio} however predict a much smaller velocity $v_{\rm
  O} \sim 1\;{\rm n m}/{\rm s}$.

\begin{figure}
\includegraphics[width=.6\columnwidth]{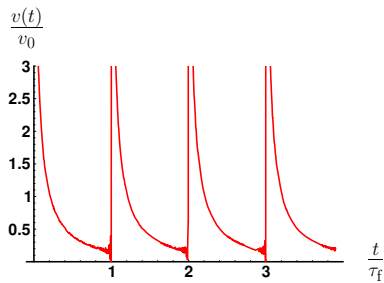}
\caption{Time evolution of the propulsive velocity for the case of
periodic particle release, corresponding to ${R}/{\sqrt{D
\tau_{\rm f}}}=1$.} \label{fig:periodic}
\end{figure}

\paragraph{Periodic Particle Release.} \label{sec:periodic}
If the reactions occur at well defined time intervals of
$\tau_{\rm f}$, then $f(\omega)=\sum_{n=-\infty}^{+\infty} e^{i
\omega n
  \tau_{\rm f}}=\frac{2 \pi}{\tau_{\rm f}} \sum_{m=-\infty}^{+\infty}
\delta(\omega-2 \pi m/\tau_{\rm f})$ in Eq.  (\ref{force-2}) and
the velocity reads
\begin{math} \displaystyle
v(t)=v_0 \; \sum_{m=-\infty}^{+\infty}\; e^{-i 2 \pi m t/\tau_{\rm
f}} {\cal A}_m \left(\frac{R}{\sqrt{D \tau_{\rm f}}}\right),
\end{math}
where
\begin{math} \displaystyle
{\cal A}_m(x)=\left[\frac{1+(1-i) \sqrt{\pi m} x}{1+(1-i)
\sqrt{\pi m} x-i \pi m x^2}\right].
\end{math}
The above expression 
can be simplified in two limiting cases. For $\frac{R}{\sqrt{D
\tau_{\rm f}}} \ll 1$, the expression in the brackets can be
approximated as $1$ and the series can be summed up to $v(t)
\simeq v_0 \sum_{n=-\infty}^{+\infty} \delta(t/\tau_{\rm f}-n)$.
This corresponds to the case where the diffusion time around the
sphere is much less than the firing time. In this case we expect
the initial inhomogeneous profile of the product particles to lead
to instantaneous impulses that are immediately screened due to
fast homogenization. In the opposite limit of $\frac{R}{\sqrt{D
\tau_{\rm f}}} \gg 1$, however, the rapid release of the product
particles maintains a stabilized inhomogeneous profile and the
velocity is approximately constant: $v(t) \simeq v_0$. The time
evolution of the velocity is plotted in Fig. \ref{fig:periodic}
for an intermediate case of $\frac{R}{\sqrt{D \tau_{\rm f}}}=1$.
We note the that the relative magnitude of the velocity
fluctuations and the average velocity is set by the ratio between
the firing time and the diffusion time.

\paragraph{Velocity Fluctuations.}     \label{sec:force-fluc}

The velocity calculated in Eq. (\ref{force-2}) should be
considered as the average of a fluctuating quantity. There are two
sources of fluctuations that need be taken into consideration for
a quantitative assessment of the velocity fluctuations, namely,
product particle density fluctuations and randomness in the
reaction that leads to the particle release.

To take account of the density fluctuations, we can go back to the
Langevin dynamics of the particles described as $\partial_t {\bf
r}_i(t)={\bf u}_i(t)$ for the velocity of the $i$-th particle,
where ${\bf u}_i(t)$ is a random noise with a distribution $P[{\bf
u}]=\exp\left[-\frac{1}{4 D} \sum_i \int dt \;{\bf
u}_i(t)^2\right]$. We can construct a stochastic density as ${\hat
\rho}({\bf r},t)=\sum_i \delta^3({\bf r}-{\bf r}_i(t))$, and show
that it satisfies Eq. (\ref{diff-eq-1}) with an additional noise
term ${\hat Q}({\bf r},t)$ added to its right hand side, whose
moments can be calculated using the above distribution as $\langle
{\hat Q}({\bf r},t) \rangle=0$ and $\langle {\hat Q}({\bf r},t)
{\hat Q}({\bf r}',t')\rangle=2 D (-\nabla^2) \delta^3({\bf r}-{\bf
r}') \delta(t-t') \rho({\bf r},t)$, where $\rho({\bf r},t)=\langle
\hat{\rho}({\bf r},t) \rangle$. To incorporate the randomness of
the reaction, we write the particle release rate as $\frac{d
N_p(t)}{d t}=\sum_n \; \delta(t-\sum_{j<n} \tau_j)$, where we have
defined a time interval $\tau_n$ between the release of the
$n+1$-th and the $n$-th particles. We further assume that the
reaction leading to the product release is a Poisson process, in
which case the probability distribution of the time intervals
between two consecutive particle release events is given by
$P(\tau)=\frac{1}{\tau_{\rm f}} \; e^{-\tau/\tau_{\rm f}}$.

Using the formulation outlined above (details of which will be
presented elsewhere \cite{ToBePub}), we can calculate the
fluctuations of the velocity about its mean value. We find the
long-time behavior of the velocity correlations as
$\overline{\langle [v(t)-v_0] [v(t')-v_0]
\rangle}\sim\frac{\ell^2}{\tau_{\rm f}} \; \delta(t-t')$, where
the over-line denotes averaging with respect to the time-interval
distribution. This result implies diffusive behavior, and we
obtain an effective diffusion coefficient
\begin{math} 
{\cal D}_{zz}\sim{\ell^2}/{\tau_{\rm f}},
\end{math}
for the sphere in the propulsion direction.
Similar results are obtained for the effective diffusion
coefficients ${\cal D}_{xx}$ and ${\cal D}_{yy}$ in the lateral
directions. We can also estimate the relative importance of the
velocity fluctuations as compared to the average velocity by
calculating $(\Delta v)^2 \equiv \overline{\langle [v(t)-v_0]^2
\rangle}$. In the long-time limit, we find
\begin{math}
\frac{(\Delta v)^2}{v_0^2}\sim {D \tau_{\rm f}}/{R^2},
\end{math}
for $R \gg a$. The velocity fluctuations are thus controlled by
the ratio between the firing time and the time it takes for the
product particles to diffuse across the sphere and homogenize
their profile. For robust propulsion, the firing time must be
considerably less than the diffusion time.

\paragraph{Effect of Hydrodynamic Flow.}  \label{sec:flow}

To make the calculations self-consistent we should also take account
of the fact that the diffusion of the product particles will be
affected by the hydrodynamic flow around the sphere. The velocity
profile of the solvent around the sphere can be shown to have a
$1/r^3$ dependence, which is characteristic of phoretic flows
\cite{Anderson-osmio}. A corresponding convective term should then be
added to Eq. (\ref{diff-eq-1}), which suggests a systematic solution
of the equations as a perturbative series in the P\'eclet number ${\rm
  Pe}=v R/D$.  Previous work has shown that the first correction term
is of the order of ${\rm Pe}^2$ for osmiophoresis
\cite{Anderson-osmio}, and ${\rm Pe} \;(\lambda_D/R)$ for
diffusiophoresis \cite{Anderson-Prieve}. These corrections are
negligible for the cases considered here (see below).

\paragraph{Reaction Kinetics.}

It is illuminating to examine more closely the conditions under
which the above simplified picture holds, taking into account the
reaction kinetics. We have a solution of substrate ${\cal S}$ at a
relatively high concentration $C_{\cal S}$ that has a natural slow
tendency to dissociate, $ {\cal S} \longrightarrow {\cal S}'+{\cal
P}$ where ${\cal S}'$ and ${\cal P}$ represent the product
particles that are assumed to exist in vanishingly small
concentrations in the bulk. This means that the entire bulk of the
system is in a non-equilibrium condition that will not relax to
equilibrium in laboratory time scale. The presence of an enzyme
${\cal E}$ catalyzes this reaction, so that it takes place
considerably faster in the vicinity of the enzyme, which leads to
an increased production of ${\cal S}'$ and ${\cal P}$ with
corresponding concentrations, $C_{{\cal S}'}$, and $C_{{\cal P}}$
respectively. There will also be a depletion of ${\cal S}$ in the
neighborhood of the enzyme. The products will slow down the
reaction by moving it towards equilibrium using the backward path
${\cal S}'+{\cal P} \longrightarrow {\cal S}$ that is inevitably
present, but one can show that it is only a small perturbation to
the reaction condition, because all the concentrations have to
match smoothly with those of the bulk that are maintained out of
equilibrium. The net propulsion of the sphere depends on all three
concentrations. If the diffusion constants (i.e. the hydrodynamic
radii) and other properties of the ${\cal S}$ and the ${\cal S}'$
particles are not that different, the sum of the two
concentrations $C_{\cal S}+C_{{\cal S}'}$ remains constant and
their contributions to the phoretic propulsion cancel~\cite{note}.
This leads to the simplified picture presented above of a
diffusive process with a localized source where $\rho=C_{{\cal
P}}$ in Eq. (\ref{diff-eq-1}).

To gain further insight on what controls the effective particle
release rate, we consider the following general multi-stage reaction pathway
\begin{equation}
{\cal S}+{\cal E}\;^{\underrightarrow{\;\;k_1\;\;}}\; {\cal
S}{\cal E}\;^{\underrightarrow{\;\;k_2\;\;}}\;{\cal P}_2 {\cal
E}+{\cal P}_1\;^{\underrightarrow{\;\;k_3\;\;}}\;{\cal P}_1+{\cal
P}_2+{\cal E},\label{reaction-general}
\end{equation}
where the two products  ${\cal P}_1$ (1st) and  ${\cal P}_2$ (2nd)
represent ${\cal S}'$ and ${\cal P}$, the order depending on which
of them (${\cal S}'$ or ${\cal P}$) is released first. The
backward reaction paths in Eq. (\ref{reaction-general}) have been
eliminated for simplicity, and they can be shown to have
negligible effect provided the nonequilibrium working condition
described above is maintained \cite{ToBePub}. Under steady-state
reaction conditions, the diffusion--reaction equations take on the
form of Eq. (\ref{diff-eq-1}), with the same rate (as a sink for
the substrate and source for the products) appearing on the right
hand side of these equations. The mean rate constant is given by
the Michaelis--Menten rule \cite{nelson}
\begin{equation}
\frac{1}{\tau_{\rm f}}=\left(\frac{k_2 k_3}{k_2+k_3}\right) \;
\frac{C_{\cal S}({\cal E})}{K_M+C_{\cal S}({\cal
E})},\label{MM-rule}
\end{equation}
where $K_M \equiv \frac{1}{k_1} \; \left(\frac{k_2
k_3}{k_2+k_3}\right)$, is the Michaelis constant and $C_{\cal
S}({\cal E})$ is the substrate concentration at the position of
the enzyme.

The concentration of the substrate at the position of the enzyme,
which we expect to be depleted in comparison to the bulk concentration
$C_0$, can be found by solving the corresponding reaction--diffusion
equation with the appropriate boundary condition. We find
\begin{eqnarray}
C_{\cal S}({\cal E})&=&\frac{1}{2}\sqrt{\left[C_0-\left(1+\frac{3
k_1 \eta a_{\cal S}}{k_{\rm B} T a_{\cal E}}\right) K_M\right]^2+4
K_M C_0}
\nonumber \\
&+&\frac{1}{2}\left[C_0-\left(1+\frac{3 k_1 \eta a_{\cal
S}}{k_{\rm B} T a_{\cal E}}\right) K_M\right],\label{C-S-E}
\end{eqnarray}
where $a_{\cal S}$ is the hydrodynamic radius of the substrate and
$a_{\cal E}$ is a typical size of the enzyme.

\paragraph{Discussion.}        \label{sec:discussion}

The molecular locomotive device will be useful if it can perform
directed motion over a distance that is large compared to its own
size.  A very important limiting factor for {directional} motion
is rotational diffusion,
whose time scale for a spherical colloid is $\tau_R=8 \pi \eta
R^3/(k_{\rm B} T)$, which implies that we can achieve more directed
motion by increasing the size of the sphere. The rotational diffusion
time turns out to be of the order of $50$ s, for $R=2 \; \mu {\rm m}$,
which means that given an optimal propulsive velocity of $v_0 \sim
0.5\, \mu \mbox{m/s}$, a locomotive of that size can travel of the
order of ten times its own size before it loses sense of its original
orientation. The effective diffusion constants derived from the
velocity fluctuations above contributes negligibly to the
orientational memory loss, giving a rotational diffusion time of the
order of $\tau_{\rm vF} \sim (R/\ell)^2 \tau_{\rm f} \simeq 1000$ s.
We also note that the typical time that the system spends in a
transient regime after a change is given by $\tau_0=R^2/D$ (see Fig.
\ref{fig:force-ini}), which is of the order of $5$ ms.  The effect of
the hydrodynamic flow can be seen to be unimportant, because ${\rm Pe}
\sim 10^{-2}$.

In conclusion, we have proposed a model design for a molecular
machine that can propel a vessel in aqueous media with a mechanism
that involves asymmetric release of reaction products. The motor
may be thought of as a diffusive equivalent of the jet engine: it
releases asymmetrically the reaction products in a viscous medium,
lets them diffuse and takes advantage of their thermodynamic
forces, instead of gaining inertial thrust by ejecting the
exhaust.

We acknowledge the support of the Royal Society.


\begin{thebibliography}{99}

\bibitem{motors}
M. Schliwa and G. Woehlke,  Nature {\bf 422}, 759 (2003).

\bibitem{armand}
F. Julicher, A. Ajdari, and J. Prost, Rev. Mod. Phys. {\bf 69},
1269 (1997); R. D. Astumian, Science {\bf 276}, 917 (1997).

\bibitem{howard} J. Howard, {\em Mechanics of Motor Proteins
and the Cytoskeleton} (Sinauer, New York, 2000).

\bibitem{mackay} see e.g. D. MacKay and R. MacKay , unpublished,
  proposing a model of trapped phosphate as a mechanism for force
  transduction in myosin:
  (\texttt{http://www.inference.phy.cam.ac.uk/mackay/dynamics/myosin/})

\bibitem{molmach}
V. Balzani {\em et al.}, Angew. Chem. Int. Ed. {\bf 39}, 3348
(2000).

\bibitem{soong}
R.K. Soong {\em et al.}, Science {\bf 290}, 1555 (2000).

\bibitem{leigh}
D.A. Leigh {\em et al.},  Nature {\bf 424}, 175 (2003).

\bibitem{Anderson-review}
J.L. Anderson, Ann. Rev. Fluid Mech. {\bf 21}, 61 (1989).

\bibitem{Lammert}
P. Lammert, J. Prost, and R. Bruinsma, J. Theor. Biol. 
{\bf 178}, 387 (1996).

\bibitem{ToBePub}
R. Golestanian, T.B. Liverpool, and A. Ajdari, to be published.

\bibitem{Derjaguin}
B.V. Derjaguin, S.S. Dukhin, A.A. Korotkova, Kolloidn. Zh. {\bf
23}, 53 (1961).

\bibitem{Anderson-osmio}
J.L. Anderson, Phys. Fluids {\bf 26}, 2871 (1983).

\bibitem{Staffeld}
P.O. Staffeld and J.A. Quinn, J. Coll. Int. Sci. {\bf 130}, 88 (1989).

\bibitem{Nardi}
J. Nardi, R. Bruinsma, and E. Sackmann, Phys. Rev. Lett. {\bf 82},
5168 (1999).

\bibitem{Anderson-Prieve}
J.L. Anderson and D.C. Prieve, Sep. Purif. Methods {\bf 13}, 67
(1984).

\bibitem{Lp}
M. Bloom, E. Evans, and O.G. Mouritsen, Q. Rev. Biophys. {\bf 24},
293 (1991).

\bibitem{Hyper}
I.S. Gradshteyn and I.M. Ryzhik, {\em Table of Integrals, Series,
and Products}, (Academic Press, New York, 2000).

\bibitem{Stryer}
L. Stryer, \textit{Biochemistry}, (Freeman, New York,
1995).

\bibitem{note}
This restriction is only imposed to simplify the presentation. If
neither of the product particles is similar to the substrate, the
net velocity of the sphere is given by the algebraic sum of the
various contributions.

\bibitem{nelson}
P. Nelson, {\em Biological Physics: Energy, Information, Life}
(Freeman, New York, 2003).



\end{thebibliography}
\end{document}